\documentclass[twocolumn,12pt]{aastex62}

\usepackage{CJK} 

\begin{document}
\begin{sloppypar}

\title{Independent Core Rotation  in Massive Filaments in Orion}

\correspondingauthor{Di Li}
\email{dili@nao.cas.cn}

\author{Xuefang Xu
\begin{CJK}{UTF8}{bsmi}
(徐雪芳)
\end{CJK}}
\affil{CAS Key Laboratory of FAST, National Astronomical Observatories, Chinese Academy of Sciences, Beijing 100101, China} 
\affil{University of Chinese Academy of Sciences, Beijing 100049, China}

\author{Di Li
\begin{CJK}{UTF8}{bsmi}
(李菂)
\end{CJK}}
\affil{CAS Key Laboratory of FAST, National Astronomical Observatories, Chinese Academy of Sciences, Beijing 100101, China} 
\affil{University of Chinese Academy of Sciences, Beijing 100049, China}
\affil{NAOC-UKZN Computational Astrophysics Centre, University of KwaZulu-Natal, Durban 4000, South Africa}

\author{Y.Sophia Dai 
\begin{CJK}{UTF8}{bsmi}
(戴昱)
\end{CJK}}
\affil{Chinese Academy of Sciences South America Center for Astronomy (CASSACA), NAOC, Beijing 100101, China} 

\author{Gary A.~Fuller}
\affiliation{Jodrell Bank Centre for Astrophysics, Department of Physics and Astronomy, The University of Manchester, Oxford Road, Manchester, M13 9PL, UK}

\author{Nannan Yue
\begin{CJK}{UTF8}{bsmi}
(岳楠楠)
\end{CJK}}
\affil{CAS Key Laboratory of FAST, National Astronomical Observatories, Chinese Academy of Sciences, Beijing 100101, China} 
\affil{University of Chinese Academy of Sciences, Beijing 100049, China}

\begin{abstract}
We present high-angular-resolution ALMA (Atacama Large Millimeter Array) 
images of N$_{2}$H$^{+}$ (1--0) that has been combined with those from the 
Nobeyama telescope  toward OMC-2 and OMC-3 filamentary regions. 
The filaments (with typical widths of $\sim$ 0.1 pc) and dense cores are resolved. 
The measured 2D velocity gradients of cores are between 1.3 and 
16.7 km\,s$^{-1}$\,pc$^{-1}$, corresponding to a specific angular momentum ($J/M$)
 between  0.0012 and 0.016 pc\,km\,s$^{-1}$. With respect to the core size $R$, 
 the specific angular momentum follows a power law $J/M \propto R^{1.52~\pm~0.14}$. 
The ratio  ($\beta$) between the rotational energy and gravitational energy ranges from 
0.00041 to 0.094, indicating insignificant support from rotation against 
gravitational collapse. We further focus on the alignment between the cores' rotational axes, 
which is defined to be perpendicular to the direction of the velocity gradient ($\theta_{G}$),
and the direction of elongation of filaments  ($\theta_{f}$) in this massive star-forming 
region.  The distribution of the angle between $\theta_{f}$ and $\theta_{G}$ was f
ound to be random, 
i.e. the cores' rotational axes have no discernible correlation with 
the elongation of their hosting filament. 
This implies that, in terms of angular momentum, 
the cores have evolved to be dynamically independent from their natal filaments.

\end{abstract}


\section{Introduction} \label{sec:intro}
High spatial and spectral resolution observations that can resolve filaments 
(with typical widths of $\sim$ 0.1\,pc) 
and cores are powerful tools for studying 
the dynamic structures of dense gas in massive star-formation regions. 
Herschel images revealed 
that, in molecular clouds, dense filaments are ubiquitous structures, 
along which dense cores are commonly found~\citep[e.g.][]{molinari2010,andre2014}. 
Dense cores, presumable site of current and future star formation can form within
or simultaneously with the filaments~\citep[e.g.][]{andre2010,arzoumanian2011,chen2015}. 
ALMA's unprecedented spectral imaging capabilities makes feasible detailed studies of the 
dynamics of cores and filaments, even in relatively distant massive star forming regions. 
In this work, we focus on the angular momentum of cores 
and its relation with respect to filaments. 

For low mass star-forming regions, earlier studies
~\citep[e.g.][]{barranco1998,caselli2002,shinnaga2004,punanova2018} 
measured the velocity gradients of dense gas at the thermal 
Jeans scale $\sim$ 0.04\,pc , such as in L1495, L1521F, TMC-1C, L1251A, PER4--7. 
For massive cores, generally more distant, the typical spatial resolution was 
$\sim$ 0.1\,pc~\citep[see e.g.][]{pirogov2003,lij2012,tatematsu2016}. 
With ALMA, we obtained high spatial resolution (3") images of N$_{2}$H$^{+}$ (1--0) 
toward filaments in OMC-2 and OMC-3. 
In conjunction with a sufficient velocity resolution of 0.11\,km\,/\,s, 
sensitive probe into the angular momentum of cores is now feasible down to 
$\sim$ 0.05\,pc scale, the thermal Jeans scale for this massive star forming regions.

The Orion Molecular Cloud (OMC), 
the closest giant molecular cloud with an OB cluster,  
is an ideal target for studying the relation between dense cores and filaments.  
OMC-2 and OMC-3 are relatively quiescent~\citep{li02, li03} 
and filamentary clouds in the OMC. 
We adopt a distance of 400\,pc for OMC-2 and OMC-3 following~\citet{nutter2007}.
~\citet{lid2013} identified 30 massive quiescent cores, 
which contain no H II region, no IRAS point sources, 
and at least 1\,pc away from the OB cluster, in OMC-2 and OMC-3. 
The core kinetic temperatures range from 13 to 31\,K. 
The majority of cores were found to be gravitationally bound and 14 cores supercritical, 
i.e., the observed thermal and non-thermal gas motion can not prevent immediate collapse 
(with a reasonable assumption of magnetic field strength $B{}\leq{}500{}\mu$ Gauss). 

Widespread and relatively easy to detect, 
N$_{2}$H$^{+}$ is a reliable tracer of cold (T $\simeq$ 10-20 K), 
dense (n(H$_2$) \textgreater 10 $\times$ 10$^{4}$ cm$^{-3}$) 
gas~\citep{bergin1997,tafalla2004,crapsi2005,tafalla2006}.
~\citet{benson1998} found that most (94\%) dense cores in their 
sample had detectable rotational transition of N$_{2}$H$^{+}$ ($J$ = 1--0). 
N$_{2}$H$^{+}$ is formed through the reaction $N_{2} + H^{+}$ and 
is mainly destroyed through reaction with CO and 
electrons~\citep{bergin1997,aikawa2001,caselli2002}.
Since the abundances of CO and electrons drop in dense gas~\citep{bergin1997}, 
the depletion timescale of N$_{2}$H$^{+}$ is longer than many other, 
particularly carbon-bound, molecules~\citep{bergin1997,aikawa2001}, 
making N$_{2}$H$^{+}$ (1--0) especially suitable for tracing dense cores and filaments.

To investigate the alignment between the core rotation and the natal filament, 
we examine the distribution of the angles between the rational axes and the 
direction of the filament elongation. 
Such an angle can potentially discriminate 
between different mechanisms of core formation.  
The gravitational fragmentation of a shock-compressed layer~\citep{whitworth1995}, 
for example,  can explain the perpendicular relation 
between the core angular momentum and filaments. 
Such configuration has indeed been found in cores with YSO 
driven outflows \citep[e.g.][]{anathpindika2008}.
In contrast, a near parallel relation 
can be explained by gravo-turbulent fragmentation~\citep[e.g.][]{banerjee2006,offner2008}.

In this letter, we characterize the orientations of the filaments in OMC-2 and OMC-3 
and measure the velocity gradients of 30 cores there  
based on ALMA N$_{2}$H$^{+}$ (1--0) images.
ALMA provided previously unattainable spatial dynamic range, 
namely the size of the longer dimension of dense structures 
divided by that of the resolution element. 
The high resolution of ALMA make feasible measuring velocity gradients of 
dense gas at the Jeans scale, in young OB-cluster-hosting regions, such as the OMC.

This letter is organized as follows. 
Observation and data are described in section~\ref{sec:data}. 
Section~\ref{sec:fit} presents our measurements of the 
filament orientation and the velocity gradients. 
In section~\ref{sec:res}, we calculate and analyze the alignment (or lack thereof) 
between  core angular momentum and its natal filament. 
We also compute the ratio between the rotational energy and gravitational energy. 
Section~\ref{sec:dis} and section~\ref{sec:sum} are the discussion 
and summary, respectively.

\section{Data} \label{sec:data}
We mapped OMC-2 and OMC-3 in N$_{2}$H$^{+}$ (1--0) with ALMA 
in November 2014 and August 2015 in band 3. 
The frequency resolution of N$_{2}$H$^{+}$ is 35\,kHz, 
which corresponds to a velocity resolution of 0.11\,km$\,s^{-1}\,$ at 93\,GHz. 
We used the 12-m main array and the Atacama Compact Array (ACA) 
to mosaic OMC-3 with 11 pointings, and OMC-2 with 7 pointings.
The baselines of the 12-m main array and the ACA observations 
are at a range of 13.6 -- 340.0\,m and 6.8 -- 87.4\,m, respectively.

The N$_{2}$H$^{+}$ (1--0) images have an angular resolution of $\sim$ 3$^{"}$, 
which  allows us to investigate the alignment between the rotation axes 
(perpendicular to the direction of velocity gradient) of cores 
and their filaments in OMC-2 and OMC-3.
This 3$^{"}$ resolution corresponds to $\sim$ 0.006\,pc, 
smaller than the characteristic scale of filaments and cores
~\citep{andre2010,molinari2010,arzoumanian2011,andre2014,chen2015}. 
This value is comparable to the Jeans scale ($\sqrt{15k_{B}T/(4{}\pi{}Gm\rho) }$) of 
$\sim$ 0.05\,pc for gas of n $\sim$ 10$^{6}$ cm$^{-3}$ at 15\,K,
which are typical conditions for the dense filaments and core envelopes in Orion.

The single dish data of N$_{2}$H$^{+}$ (1--0) were taken from the NRO 
(Nobeyama Radio Observatory) Star Formation Legacy Project~\citep{nakamura2019}. 
The Nobeyama N$_{2}$H$^{+}$ observations were executed with 
the newly-developed 100\,GHz-band 4-beam dual-polarization receiver (FOREST) 
during January and March 2017. 
The Nobeyama velocity resolution is 0.1\,km$\,s^{-1}\,$. 
The system temperature and noise level are at a range of 150 -- 200\,K 
and 0.26 -- 0.30\,K, respectively.

\begin{figure*}[!htb]
\centering
\includegraphics[width=0.80\textwidth]{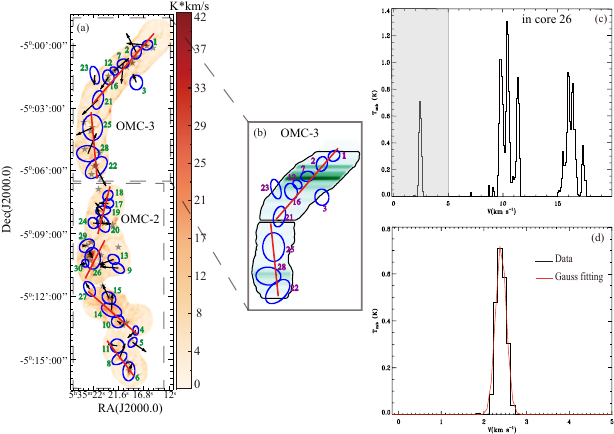}
\caption{(a) 30 cores overlaid with the integrated intensity of N$_{2}$H$^{+}$ (1--0). 
The blue ellipses are the size and location of the 30 cores. 
The black arrows with varying length represent the velocity gradients, 
as measured in section~\ref{sec:gra}. 
The lengths of black arrows are the values of the measured velocity gradients.
The red solid lines are the fitted filaments (see section~\ref{sec:fila}). 
The gray stars are YSOs (Young Stellar Objects). 
(b) Zoomed in view of the fitted filaments and their cores in OMC-3. 
The background is the autocorrelation map of 
the N$_{2}$H$^{+}$ integrated intensity map.
The contour marks the 20\% of the maximum peak in the autocorrelation map.
(c) The brightest N$_{2}$H$^{+}$ peak (in core 26) in OMC-2 and OMC-3.
Though N$_{2}$H$^{+}$ (1--0) has hyperfine components, 
the shaded region refers to a single isolated gaussian component 
due to the N$_{2}$H$^{+}$ (1--0) transition.
(d) Gaussian fit to the line in the shaded region of (c). \label{fig:direc}} 
\end{figure*}

We combined the N$_{2}$H$^{+}$ data from ALMA and Nobeyama 
to recover missing fluxes in the interferometric data. 
The combination was performed using CASA 
(Common Astronomy Software Applications)~\citep{mcmullin2007}. 
The integrated intensity of the combined N$_{2}$H$^{+}$ (1--0) images 
and the overlaid cores, labelled 1 to 30, are shown in Figure~\ref{fig:direc}(a).

\section{Fitting filament and velocity gradient}  \label{sec:fit}
\subsection{Orientations of the filaments} \label{sec:fila}
In this section, we focus on the large-scale filamentary structures ($\geq${}0.3\,pc)  
and fit them following the procedures described in~\citet{lihb2013}. 
For the N$_{2}$H$^{+}$ integrated density map,  
the long-axis direction of the autocorrelation function 
is defined as the filament orientation ($\theta_{f}$).
The fitted filaments in OMC-2 and OMC-3 are shown 
as the red solid lines in Figure~\ref{fig:direc}(a).

The main steps of fitting filaments to obtain $\theta_{f}$ 
are described as follows. 
We first apply a two--dimensional Fourier transform 
to the N$_{2}$H$^{+}$ integrated intensity map, 
which is then multiplied by its complex conjugate. 
The inverse transform of the product yields 
the autocorrelation function. 
For a given contour of this autocorrelation map, 
the long axis direction ($\theta_{f}$) can be obtained by performing 
a least-squares fit to the pixel positions located within the contour. 
The fitting of $\theta_{f}$ is robust with respect to the choice of
contour levels, as only $< 5^{\circ}$ variation was found among different trials. 
In subsequent calculations,  the contour at 20\% of the peak value was adopted. 
Figure~\ref{fig:direc}(b) displays the results from steps above in OMC-3.

\subsection{Fitting spectral lines and velocity gradient} \label{sec:gra}
Though N$_{2}$H$^{+}$ (1--0) has 7, closely-spaced hyperfine components, 
there is an isolated one (the shaded region in Figure~\ref{fig:direc}(c)).  
For measuring the velocity gradient, 
we compared fitting all hyperfine components 
and fitting just the isolated one with a single Gaussian.
No significant difference was found. 
Therefore, we relied on the single-Gaussian fit in all subsequent calculations.

To securely measure the velocity gradient, we require 
the spectral lines to have peak intensity greater than 
three times of the Root Mean Square (rms) noise 
and fitting errors in the cores. 
We then fit the velocity field as 
$v_{LSR} = v_0 + a \Delta \alpha + b \Delta \delta$, 
and measure the velocity gradients following the method 
described in~\citet{goodman1993}. 
$\Delta \alpha$ and $\Delta \delta$ are the offsets 
from the center position 
(5$^{h}$:35$^{m}$:21.0$^{s}$, -5$^{\circ}$:00$^{'}$:00$^{''}$) 
of the combined N$_{2}$H$^{+}$ images 
in the right ascension and declination in radians, respectively.  
a and b are the projections of the gradient per radian 
onto the $\alpha$ and $\delta$ axes, respectively.
The magnitude of the velocity gradient is given by
\begin{equation}
{\cal G}= |\mathbf\nabla
v_\mathrm{LSR}|=\frac{(a^2+b^2)^{1/2}}{D},
\end{equation}
where $D$ denotes the distance to the cores.
The gradient direction is given by
\begin{equation}
\theta_{G}= \tan{}\frac{a}{b}.
\end{equation}
The fitted $\cal{}G$ and $\theta_{G}$ are listed in Table~\ref{tab:gar}. 
The black arrows (Figure~\ref{fig:direc}(a)) with varying length 
illustrate the velocity gradients.

\begin{deluxetable*}{ccccccccccccc}
\tablecaption{Dynamic Parameters of Orion Cores \label{tab:gar}}
\tablewidth{600pt}
\tabletypesize{\tiny}
\tablehead{
\colhead{Core} & \colhead{R.A.} & \colhead{Decl.} & \colhead{Mass} &
\colhead{Mass$_{vir}$} & \colhead{Major} &  \colhead{Minor} &
 \colhead{Major Axis Orientation} & \colhead{$\cal G$} & \colhead{$\theta_{G}$} & 
\colhead{J/M} & \colhead{$\beta$} & 
\colhead{ $\vert{}\theta_{f} - \theta_{G}{}\vert$} \\ 
\colhead{} & \colhead{(J2000)} & \colhead{(J2000)} & \colhead{(M$_\sun$)} & 
\colhead{(M$_\sun$)} & \colhead{"} & \colhead{"} &  \colhead{(deg E of N)} &  \colhead{(km$\,s^{-1}\,pc^{-1}$)} &  \colhead{ (deg E of N)} & 
\colhead{(pc\,km$\,s^{-1}$)} & \colhead{} & \colhead{(deg)} 
} 
\startdata
1  & 5:35:16.1 & -5:00:00 & 12.4 & 6.6 & 16.0 & 13.0 &   3 &  2.2 $\pm$ 0.3 &   47 $\pm$ 6 & 1.2 $\times$10$^{-3}$ $\pm$ 1.7$\times$10$^{-4}$  & 1.7$\times$10$^{-3}$ $\pm$ 3.3$\times$10$^{-5}$ &  81 $\pm$ 12\\   
2  & 5:35:18.2 & -5:00:21 & 25.9 & 5.2 & 19.0 & 14.0 & 12  &  1.6 $\pm$ 0.3 &   59 $\pm$ 8 & 1.9 $\times$10$^{-3}$ $\pm$ 1.8$\times$10$^{-4}$  & 1.8$\times$10$^{-3}$ $\pm$ 1.5$\times$10$^{-5}$ &  38 $\pm$ 16\\   
3  & 5:35:18.2 & -5:01:47 &  3.1 & 3.0 & 21.0 & 18.0 & 52  & 10.3 $\pm$ 2.6 &   29 $\pm$ 9 & 9.5 $\times$10$^{-3}$ $\pm$ 1.6$\times$10$^{-3}$  & 3.2$\times$10$^{-2}$ $\pm$ 8.7$\times$10$^{-4}$ &  63 $\pm$  6\\   
4  & 5:35:18.3 & -5:13:38 &  0.3 & 0.4 & 13.0 &  8.0 & 35 &  9.1 $\pm$ 1.1 &  -86 $\pm$ 3 & 3.8 $\times$10$^{-3}$ $\pm$ 2.2$\times$10$^{-4}$  & 7.8$\times$10$^{-2}$ $\pm$ 2.8$\times$10$^{-4}$ &  79 $\pm$ 12\\   
5  & 5:35:18.9 & -5:14:12 &  0.5 & 1.2 & 16.0 & 10.0 & 55 &  5.1 $\pm$ 0.8 &   35 $\pm$ 6 & 1.9 $\times$10$^{-3}$ $\pm$ 2.8$\times$10$^{-4}$  & 1.1$\times$10$^{-2}$ $\pm$ 2.5$\times$10$^{-4}$ &  22 $\pm$ 3 \\   
6  & 5:35:19.6 & -5:15:35 &  5.8 & 5.7 & 27.0 & 17.0 &  3 & 11.1 $\pm$ 0.3 &  139 $\pm$ 1 & 1.1 $\times$10$^{-2}$ $\pm$ 2.7$\times$10$^{-4}$  & 1.8$\times$10$^{-2}$ $\pm$ 1.1$\times$10$^{-5}$ &  54 $\pm$ 19\\   
7  & 5:35:20.7 & -5:00:53 & 30.0 & 1.8 & 17.0 & 13.0 & 14 &  2.6 $\pm$ 0.2 &  -14 $\pm$ 3 & 1.3 $\times$10$^{-3}$ $\pm$ 1.3$\times$10$^{-4}$  & 8.0$\times$10$^{-4}$ $\pm$ 8.0$\times$10$^{-6}$ &  19 $\pm$ 22\\   
8  & 5:35:21.4 & -5:14:58 &  4.0 & 2.1 & 24.0 & 14.0 & 26 &  5.6 $\pm$ 0.3 &  124 $\pm$ 2 & 3.4 $\times$10$^{-3}$ $\pm$ 2.3$\times$10$^{-4}$  & 2.9$\times$10$^{-2}$ $\pm$ 1.3$\times$10$^{-4}$ &  69 $\pm$ 6 \\   
9  & 5:35:21.6 & -5:10:39 &  1.6 & 1.4 & 19.0 & 12.0 & 47 &  4.7 $\pm$ 0.4 &    3 $\pm$ 3 & 1.7 $\times$10$^{-3}$ $\pm$ 2.1$\times$10$^{-4}$  & 2.4$\times$10$^{-2}$ $\pm$ 3.6$\times$10$^{-4}$ &   6 $\pm$ 9 \\   
10 & 5:35:21.7 & -5:13:12 &  9.2 & 6.8 & 18.0 & 18.0 & 19 & 16.2 $\pm$ 0.4 &   70 $\pm$ 1 & 8.3 $\times$10$^{-3}$ $\pm$ 2.6$\times$10$^{-4}$  & 9.5$\times$10$^{-2}$ $\pm$ 9.4$\times$10$^{-5}$ &  55 $\pm$ 8 \\   
11 & 5:35:21.8 & -5:14:22 &  5.7 & 4.0 & 20.0 & 19.0 & 55 & 10.2 $\pm$ 0.2 &   82 $\pm$ 1 & 8.5 $\times$10$^{-3}$ $\pm$ 1.9$\times$10$^{-4}$  & 1.4$\times$10$^{-2}$ $\pm$ 7.6$\times$10$^{-5}$ &  69 $\pm$ 22\\   
12 & 5:35:22.4 & -5:01:14 & 15.6 & 1.2 & 13.0 & 10.0 &  1 &  3.1 $\pm$ 0.7 &  136 $\pm$ 9 & 6.0 $\times$10$^{-4}$ $\pm$ 2.0$\times$10$^{-4}$  & 4.1$\times$10$^{-4}$ $\pm$ 4.6$\times$10$^{-5}$ &  11 $\pm$ 16\\   
13 & 5:35:22.5 & -5:10:14 &  2.5 & 5.4 & 17.0 & 13.0 & 37 &  5.9 $\pm$ 0.3 &  -15 $\pm$ 2 & 3.2 $\times$10$^{-3}$ $\pm$ 1.7$\times$10$^{-4}$  & 5.6$\times$10$^{-2}$ $\pm$ 1.5$\times$10$^{-4}$ &  12 $\pm$ 10\\   
14 & 5:35:22.9 & -5:12:40 & 26.3 & 6.7 & 30.0 & 18.0 & 37 &  5.2 $\pm$ 0.4 &  121 $\pm$ 3 & 9.3 $\times$10$^{-3}$ $\pm$ 3.4$\times$10$^{-4}$  & 2.6$\times$10$^{-2}$ $\pm$ 3.5$\times$10$^{-5}$ &  74 $\pm$ 21\\   
15 & 5:35:23.4 & -5:12:05 & 12.8 & 3.2 & 21.0 & 16.0 &  2 &  6.2 $\pm$ 0.6 &   61 $\pm$ 4 & 2.6 $\times$10$^{-3}$ $\pm$ 3.1$\times$10$^{-4}$  & 5.8$\times$10$^{-3}$ $\pm$ 8.3$\times$10$^{-5}$ &  46 $\pm$ 5 \\   
16 & 5:35:23.5 & -5:01:32 & 23.1 & 5.7 & 21.0 & 16.0 &  3 &  6.6 $\pm$ 0.4 &  -24 $\pm$ 2 & 4.0 $\times$10$^{-3}$ $\pm$ 2.5$\times$10$^{-4}$  & 7.6$\times$10$^{-3}$ $\pm$ 2.9$\times$10$^{-5}$ &   9 $\pm$ 13\\   
17 & 5:35:23.5 & -5:07:34 &  7.6 & 3.0 & 15.0 & 13.0 & 39 & 10.0 $\pm$ 0.5 &   28 $\pm$ 2 & 4.6 $\times$10$^{-3}$ $\pm$ 2.2$\times$10$^{-4}$  & 4.2$\times$10$^{-2}$ $\pm$ 9.2$\times$10$^{-5}$ &  42 $\pm$ 20\\   
18 & 5:35:23.6 & -5:07:11 &  6.1 & 3.4 & 16.0 & 13.0 & 32 & 16.6 $\pm$ 0.5 &  -55 $\pm$ 1 & 8.5 $\times$10$^{-3}$ $\pm$ 2.3$\times$10$^{-4}$  & 1.7$\times$10$^{-2}$ $\pm$ 1.2$\times$10$^{-4}$ &   9 $\pm$ 4 \\   
19 & 5:35:24.5 & -5:07:54 & 11.7 & 3.1 & 21.0 & 13.0 & 17 &  9.1 $\pm$ 0.5 &   70 $\pm$ 3 & 6.0 $\times$10$^{-3}$ $\pm$ 3.2$\times$10$^{-4}$  & 3.5$\times$10$^{-2}$ $\pm$ 9.7$\times$10$^{-5}$ &  84 $\pm$ 2 \\   
20 & 5:35:24.5 & -5:08:32 &  9.7 & 3.7 & 26.0 & 16.0 &  3 &  4.2 $\pm$ 0.2 &   80 $\pm$ 2 & 2.4 $\times$10$^{-3}$ $\pm$ 1.9$\times$10$^{-4}$  & 5.7$\times$10$^{-3}$ $\pm$ 3.6$\times$10$^{-5}$ &  86 $\pm$ 7 \\   
21 & 5:35:25.5 & -5:02:37 &  6.8 & 3.5 & 27.0 & 15.0 & 63 &  1.6 $\pm$ 0.2 &  -10 $\pm$ 4 & 1.7 $\times$10$^{-3}$ $\pm$ 2.1$\times$10$^{-4}$  & 4.0$\times$10$^{-3}$ $\pm$ 5.9$\times$10$^{-5}$ &  23 $\pm$ 11\\   
22 & 5:35:25.8 & -5:05:51 & 13.0 & 4.3 & 38.0 & 22.0 &  7 &  2.0 $\pm$ 0.1 & -118 $\pm$ 3 & 5.5 $\times$10$^{-3}$ $\pm$ 2.1$\times$10$^{-4}$  & 1.5$\times$10$^{-2}$ $\pm$ 2.1$\times$10$^{-5}$ &  66 $\pm$ 6 \\   
23 & 5:35:26.1 & -5:01:26 &  3.8 & 0.7 & 26.0 & 12.0 & 39 &  9.8 $\pm$ 1.0 &  -71 $\pm$ 5 & 8.8 $\times$10$^{-3}$ $\pm$ 5.1$\times$10$^{-4}$  & 1.8$\times$10$^{-2}$ $\pm$ 7.0$\times$10$^{-4}$ &  37 $\pm$ 10\\   
24 & 5:35:26.4 & -5:08:30 &  3.8 & 5.9 & 16.0 & 14.0 & 19 &  4.4 $\pm$ 0.6 & -107 $\pm$ 7 & 2.9 $\times$10$^{-3}$ $\pm$ 2.8$\times$10$^{-4}$  & 3.2$\times$10$^{-2}$ $\pm$ 2.9$\times$10$^{-4}$ &  87 $\pm$ 7 \\   
25 & 5:35:26.5 & -5:03:56 & 13.2 &12.8 & 37.0 & 28.0 &  2 &  4.1 $\pm$ 0.1 &  -70 $\pm$ 1 & 6.8 $\times$10$^{-3}$ $\pm$ 2.0$\times$10$^{-4}$  & 2.1$\times$10$^{-2}$ $\pm$ 1.9$\times$10$^{-5}$ &  55 $\pm$ 4 \\   
26 & 5:35:26.5 & -5:10:11 & 33.5 & 8.3 & 31.0 & 22.0 & 10 & 10.8 $\pm$ 0.2 &   14 $\pm$ 1 & 1.6 $\times$10$^{-2}$ $\pm$ 2.8$\times$10$^{-4}$  & 5.8$\times$10$^{-2}$ $\pm$ 1.8$\times$10$^{-5}$ &  17 $\pm$ 15\\   
27 & 5:35:27.1 & -5:11:39 &  2.3 & 4.5 & 23.0 & 14.0 & 38 & 11.9 $\pm$ 0.7 &   37 $\pm$ 2 & 2.5 $\times$10$^{-3}$ $\pm$ 5.1$\times$10$^{-4}$  & 2.7$\times$10$^{-2}$ $\pm$ 1.2$\times$10$^{-5}$ &  22 $\pm$ 7 \\   
28 & 5:35:27.4 & -5:05:11 & 20.1 & 3.2 & 32.0 & 23.0 & 16 &  1.3 $\pm$ 0.1 &  108 $\pm$ 3 & 2.5 $\times$10$^{-3}$ $\pm$ 1.5$\times$10$^{-4}$  & 2.2$\times$10$^{-4}$ $\pm$ 8.1$\times$10$^{-6}$ &  79 $\pm$ 4 \\   
29 & 5:35:27.6 & -5:09:35 & 12.9 & 12.8& 23.0 & 17.0 &  1 & 16.7 $\pm$ 0.5 &   95 $\pm$ 1 & 9.9 $\times$10$^{-3}$ $\pm$ 3.6$\times$10$^{-4}$  & 7.6$\times$10$^{-2}$ $\pm$ 1.0$\times$10$^{-4}$ &  41 $\pm$ 21\\   
30 & 5:35:27.9 & -5:10:25 &  6.4 & 3.3 & 11.0 & 10.0 & 16 & 12.8 $\pm$ 1.1 &  -41 $\pm$ 4 & 3.9 $\times$10$^{-3}$ $\pm$ 2.4$\times$10$^{-4}$  & 5.0$\times$10$^{-3}$ $\pm$ 1.8$\times$10$^{-4}$ &  38 $\pm$ 8 \\   
\enddata
 \tablecomments{  
 The values of Mass and Mass$_{vir}$ are from~\citet{lid2013}.
 Major and Minor axes are from~\citet{nutter2007}.
 Major axis orientation is the major axis direction of cores.}
\end{deluxetable*}

\section{Results} \label{sec:res}
In this section, we calculate and analyze 
the distribution of the angles between filaments and 
velocity gradients, based one the fitted angular differences
$\vert{}\theta_{f} - \theta_{G}{}\vert$.
We also derive the specific angular momentum $J/M$ 
and the ratio $\beta$ between the rotational energy 
and gravitational energy, based on the measured velocity gradients. 

\subsection{The distribution of $\vert{}\theta_{f} - \theta_{G}{}\vert$}
\begin{figure*}[!htb]
\centering
\includegraphics[width=0.80\textwidth]{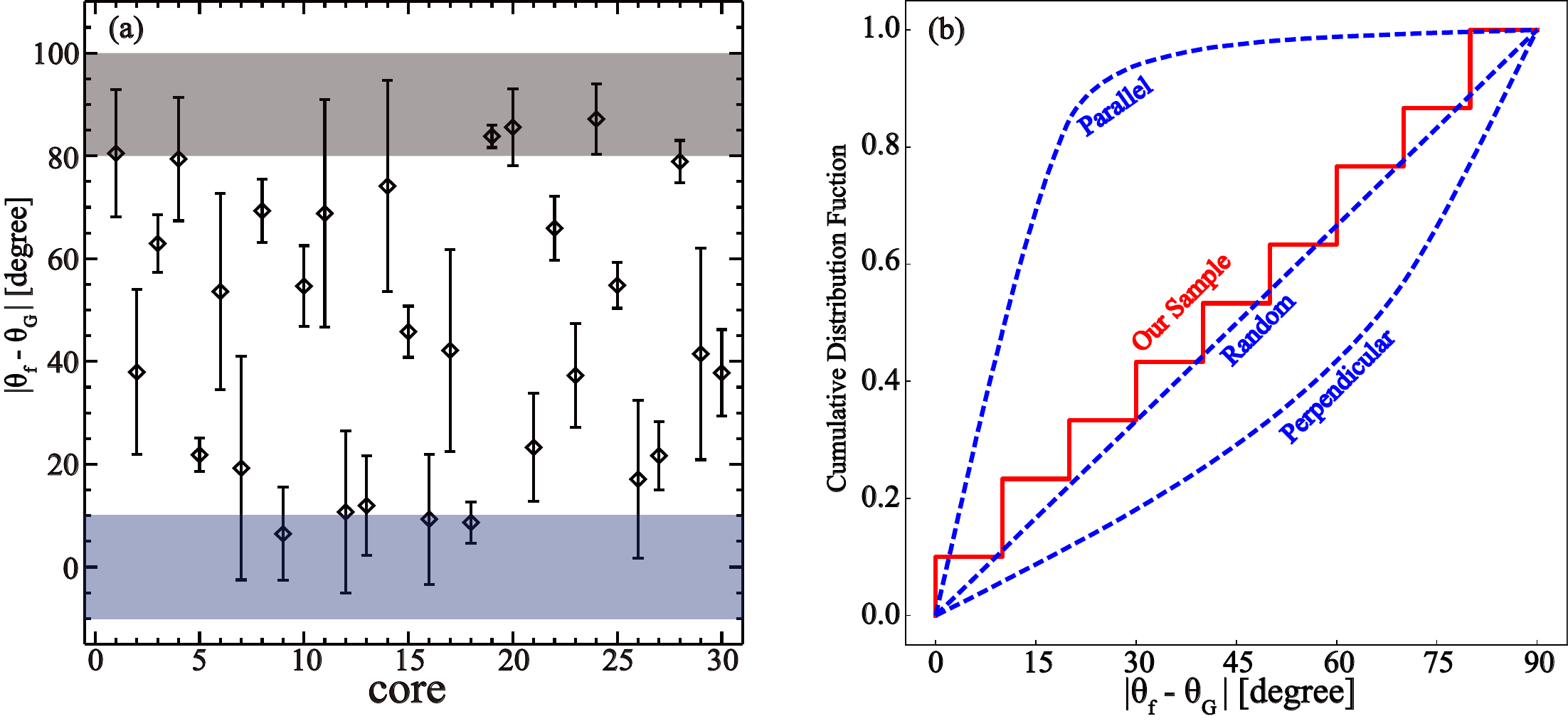}
\caption{(a) $\vert{}\theta_{f} - \theta_{G}{}\vert$ for each of the 30 cores. 
The gray strip corresponds to the perpendicular directions, 
i.e. 90$^{\circ}$ $\pm$10$^{\circ}$.  
The dusty blue strip corresponds to the parallel directions, 
i.e. 0$^{\circ}$ $\pm$10$^{\circ}$. 
(b) The cumulative distribution function of 
$\vert \theta_{f} - \theta_{G} \vert$ 
and the projected $\theta_{3D}$. 
The red step line is the 
$\vert \theta_{f} - \theta_{G} \vert$ of our sample. 
The three blue dashed lines are results
from the Monte Carlo simulation. \label{fig:ks}}
\end{figure*}

To quantify the alignment between the velocity gradients of cores 
and their filaments, $\vert{}\theta_{f} - \theta_{G}{}\vert$ 
is derived as
\begin{equation}
\vert{}\theta_{f} - \theta_{G}{}\vert = MIN\{{}\vert{}\theta_{f} - \theta_{G}{}\vert, 180-\vert{}\theta_{f} - \theta_{G}{}\vert{}\}\label{equ1},
\end{equation}
where $\theta_{f}$ and $\theta_{G}$ are measured 
clockwise from the East,
and `MIN' refers to the minimum angular difference 
between $\theta_{f}$ and $\theta_{G}$. 
The derived $\vert{}\theta_{f} - \theta_{G}{}\vert$ values are listed 
in Table~\ref{tab:gar} and plotted in Figure~\ref{fig:ks}(a). 
We find that the gradient directions of 7 cores ($\sim$23\%)  
are essentially parallel to their filament orientations 
(0$^{\circ}$ $\pm$10$^{\circ}$, 
masked as the dusty blue strip in Figure~\ref{fig:ks}(a)), 
while the gradient directions of 8 cores ($\sim$27\%) 
are essentially perpendicular to their filaments 
(90$^{\circ}$ $\pm$10$^{\circ}$, 
the gray strip in Figure~\ref{fig:ks}(a)).

To investigate the distribution of 
$\vert{}\theta_{f} - \theta_{G}{}\vert$,  
Monte Carlo simulations are performed 
in three-dimensional (3D) space. 
We generate two random unit vectors 
within a unit sphere in 3D, 
and measure the angle ($\theta_{3D}$) between the two vectors. 
10$^{6}$ pairs of unit vectors are generated 
to produce 10$^{6}$ angles of $\theta_{3D}$,
constrained to a range of 0$^{\circ}$ -- 90$^{\circ}$. 
If $\theta_{3D}$ is larger than 90$^{\circ}$, 
the 180$^{\circ}$ - $\theta_{3D}$ values are adopted (Equation~\ref{equ1}). 
For $\theta_{3D}$ in the range of 0$^{\circ}$ -- 20$^{\circ}$, 
we define it parallel,
while 20$^{\circ}$ -- 70$^{\circ}$ is defined random, 
and 70$^{\circ}$ -- 90$^{\circ}$ perpendicular. 
Then we project the angles of $\theta_{3D}$ 
onto a two-dimensional (2D) space.
Figure~\ref{fig:ks}(b) plots the cumulative distribution function of 
our $\vert{}\theta_{f} - \theta_{G}{}\vert$ 
and the projected $\theta_{3D}$.  
The details of the Monte Carlo simulations can be found 
in Appendix A of~\citet{stephens2017}.
The $p$--values of the Anderson -- Darling (AD) test 
are used to indicate 
whether two distributions are consistent. 
$p$--values near 1 imply that the two distributions 
are likely consistent, 
while $p$--values near 0 imply that they are not consistent. 
For the distribution of our $\vert \theta_{f} - \theta_{G}{}\vert$ 
and the simulated random distribution, 
the AD test gives a $p$--value of 0.97, consistent of being random. 
This result indicates no correlation between the gradient direction 
of a core and its filament orientation.
This is consistent with the scenario that the rotation axis of a core 
is independent of the orientation of the large-scale filament 
in which it resides. 

\subsection{J/M and $\beta$ calculations}
\begin{figure*}[!htb]
\centering
\includegraphics[width=0.80\textwidth]{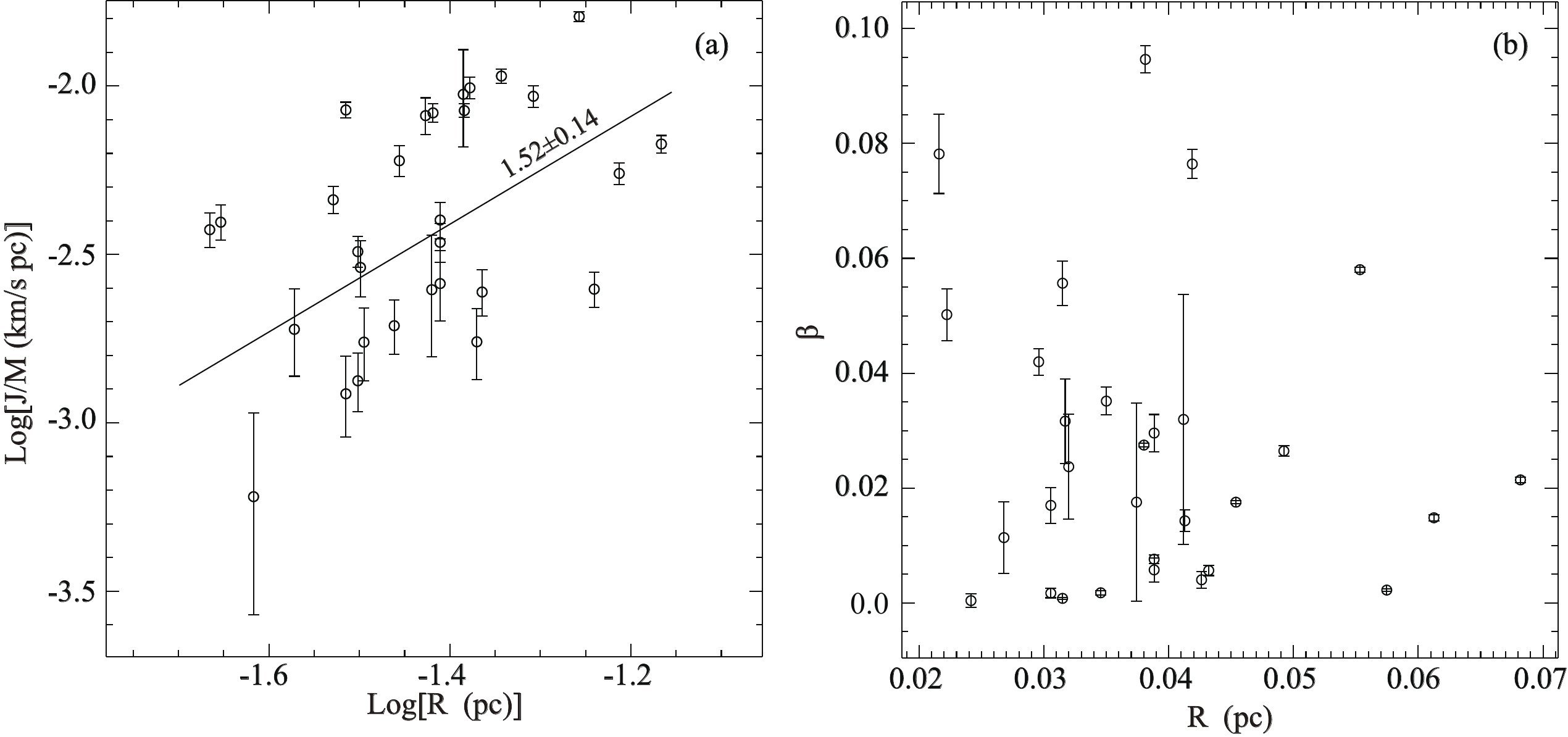}
\caption{The correlation between core size $R$ 
and (a) specific angular momentum $J/M$ and (b) $\beta$: ratio between rotational and gravitational energy. 
The best-fit power-law relation between $J/M$ and $R$, 
as well as corresponding coefficients, 
are labelled in (a).\label{fig:dyn}}
\end{figure*}

The $J/M$ and $\beta$ calculations were performed 
under the assumption that cores have a power-law profile density 
($\rho \propto R^{-1.6}$)~\citep{bonnor1956}. 
We used the following equations to calculate $J/M$ and $\beta$: 
\begin{equation}
J/M =  0.59{}{\cal G} R^{2}, 
\end{equation}
and
\begin{equation}
\beta = \frac{0.34{}{\cal G}^{2}R^{3}}{GM}. 
\end{equation}
$M$ is the mass within a radius $R$. 
The calculated $J/M$ and $\beta$ are listed in Table~\ref{tab:gar}, 
and plotted in Figure~\ref{fig:dyn}(a) and Figure~\ref{fig:dyn}(b), 
respectively. 
We find that $J/M$ increases with increasing $R$, 
following a power-law $J/M~\propto~R^{1.52~\pm~0.14}$. 
The slope is consistent with the index value of 1.6 found by
~\citet{goodman1993} for a sample of dark clouds. 
$\beta$ ranges from 0.00041 to 0.094, a no significant correlation with $R$. 
As investigated in previous studies
~\citep{goldsmith1985,goodman1993,caselli2002,curtis2011,tatematsu2016}, 
these small values of $\beta$ indicate that rotation alone is not enough to 
support the core from collapsing due to gravity.
 
\section{Discussion}\label{sec:dis}
Magnetic fields play an important role during the collapse and fragmentation 
of massive molecular clumps, as well as the formation of dense cores.
~\citet{zhang2014} found that the magnetic fields at dense core scales 
are either aligned with  
or perpendicular to the parsec-scale magnetic fields. 
One may reasonably expect such bimodality for the
angles between core rotation and filament elongation.
~\citet{anathpindika2008} reported that 
the rotational axes of prestellar cores are perpendicular to their filaments.
~\citet{stephens2017} showed that a mix of parallel and perpendicular 
angles exist between the rotation axes of protostellar cores and their filaments. 
In our sample, however, we find no correlation between the rotational direction 
and the natal filament's elongation. 
The distribution is found to be consistent with being random. 
Similar results were also seen in earlier observations~\citep[e.g.][]{heyer1988,davis2009,lee2016,offner2016,punanova2018}. 

Considerable difference exist among the analyses mentioned above, in terms of sample sizes, definition of filament and cores, and the fitting of rotation and elongation.
~\citet{anathpindika2008} and~\citet{stephens2017} regarded 
the outflow orientations as the rotation axes of prestellar/protostellar cores,
while the directions of velocity gradients trace the rotation axes 
of cores in~\citet{punanova2018}, as well as in our sample.
~\citet{stephens2017} identified filaments using 
FILFINDER~\citep{koch2015} and SExtractor~\citep{bergin1996},
while some studies~\citep[e.g.][]{arzoumanian2018,bresnahan2018,gong2018,tanimura2019} 
employed DisPerSE~\citep{sousbie2011} to trace filaments.    
Although our experience seems to suggest that these technical
differences are unlikely to explain the diversity in the relative orientations 
between rotation and filament, high spectral and spatial images of
 a much larger sample are needed to tackle the issue systematically. 

\section{Summary}\label{sec:sum}
We mapped N$_{2}$H$^{+}$ (1--0) in OMC-2 and OMC-3 
with both ALMA and Nobeyama at 93\,GHz. 
The combined single dish and interferometric data offer a rare opportunity 
to study the relative orientation between the rotational axes of 
the massive quiescent cores and the elongation of their natal filaments, down to the thermal Jeans scale of dense gas in a massive star forming region. 
Our results are summarized as the following.

1. The angle ($\vert{}\theta_{f} - \theta_{G}{}\vert$) between the orientation of the filaments ($\theta_{f}$) and the direction of core velocity gradient $\theta_{f}$ was found to be random, based on a Monte-Carlo simulation. The core rotation seems to have disengaged itself from its natal filament, and by association, the large-scale magnetic field.

2. The measured velocity gradients of 30 cores 
range from 1.3 to 16.7 km\,s$^{-1}$\,pc$^{-1}$. 
The measured specific angular momentum ($J/M$) 
ranges from 0.0012 to 0.016 pc\,km\,s$^{-1}$. 
A power-law scaling was found between the specific angular momentum and 
the core size as  $J/M~\propto~R^{1.52~\pm~0.14}$. 

3. The ratio $\beta$ between the rotational and gravitational energy 
ranges from 0.00041 to 0.094, 
indicating that rotation cannot stop gravitational collapse in these dense cores.

This work is supported by the National Natural Science Foundation of China grant No.\ 11988101, No.\ 11725313, No.\ 11721303, the International Partnership Program of Chinese Academy of Sciences grant No.\ 114A11KYSB20160008.
This paper makes use of the following ALMA data: ADS/JAO.ALMA\#2013.1.00662.S. 
ALMA is a partnership of ESO (representing its member states), NSF (USA) and NINS (Japan), together with NRC (Canada), MOST and ASIAA (Taiwan), and KASI (Republic of Korea), in cooperation with the Republic of Chile. The Joint ALMA Observatory is operated by ESO, AUI/NRAO and NAOJ.

\end{sloppypar}
\end{document}